\documentclass{aastex}
\begin{document}
\title{The Steady Spin Down Rate of 4U 1907+09      
 }

   \author{Altan Baykal $^{1}$, 
          \,{C}a\u{g}da\c{s} \.{I}nam $^{1}$,
          M. Ali Alpar $^{2}$,\\
          Jean in't Zand $^{3}$,
          Tod Strohmayer $^{4}$}
\affil{ 
 $^{1}$ Physics Department, Middle East Technical University,
  Ankara 06531, Turkey,\\
 $^{2}$ Faculty of Engineering and Natural Sciences, 
 Sabanc{\i} University, 81474, Istanbul, Turkey\\
 $^{3}$ Space Research Organization 
 Sorbonnelaan 2, 3584 CA Utrecht, Netherlands, \\
 $^{4}$ Laboratory for High Energy Astrophysics NASA/GSFC
  Greenbelt, Maryland 20771 USA }

   \begin{abstract}

Using X-ray data from the Rossi X-Ray Timing Explorer (RXTE), we report
the pulse timing results of the 
accretion powered high mass X-ray binary (HMXRB) pulsar
4U 1907+09 covering a time span of almost  
two years. We measured three new pulse periods 
in addition to the previously measured four pulse periods. We are
able to connect pulse arrival times in phase for more than a year. 
The source has been spinning down almost at a constant rate with
 a spin down rate of 
$\dot \nu$ = (-3.54 $\pm  0.02) \times 10^{-14}$ Hz s$^{-1}$ for more than  
 15 years. Residuals of pulse arrival times yield a very low
level of random walk noise strengths
$\sim 2 \times 10^{-20}$ rad$^{2}$ sec$^{-3}$ 
on a time scale of 383 days, which is four decades 
lower than that of the HMXRB pulsar Vela X-1. 
The noise strength is only a factor of 5 greater than
that of the low mass X-ray binary pulsar (LMXRB)
4U 1626-67.
The low level of the timing noise and the very stable spin down rate 
of 4U 1907+09 makes this source unique among the HMXRBs, 
providing another example, in addition to 4U 1626-67, 
of long term quiet spin down from an accreting source.
These examples show that the extended quiet spin down episodes 
observed in the anomalous X-ray pulsars (AXPs) pulsars
1RXS J170849.0-400910 and 1E 2259+586 
do not necessarly imply that these sources are not accreting 
pulsars.


\end{abstract} 
\section{Introduction}
4U 1907+09 is an accretion powered X-ray binary pulsar
which is accreting plasma from a blue supergiant companion star. 
It was discovered as an X-ray source by Giacconi et al. (1971) 
and has been studied using instruments on board  Ariel V 
(Marshall $\&$ Ricketts 1980), Tenma (Makishima et al., 1984), 
EXOSAT (Cook $\&$ Page 1987), Ginga (Makishima $\&$ Mihara 1992, Mihara 1995), 
and RXTE (In 't Zand, Strohmayer $\&$ Baykal 1997, In 't Zand, Baykal $\&$ 
Strohmayer 1998, In 't Zand, Strohmayer $\&$ Baykal 1998).
Marshall $\&$ Ricketts (1980) first determined the orbital period of the 
binary at 8.38 days by analysing the data taken  
between 1974 and 1980 from a survey instrument on board Ariel V. 
Folding the light curve of these data, they found two flares,
a primary and a secondary, each occurring at the same orbital phase. 
Subsequent Tenma observations of this source have shown a pulse period at 
437.5 sec (Maksihima et al., 1984).
 Later EXOSAT (Cook $\&$ Page 1987) and recent RXTE obsevations
(In 't Zand, Baykal $\&$ Strohmayer 1998,
 In 't Zand, Strohmayer $\&$ Baykal 1998) 
have shown that these flares are locked to orbital phases  
separated by half an orbital period. Makishima et al., (1984) and 
Cook $\&$ Page (1987) suggested that the two flares are due to an equatorial 
disk-like envelope around a companion star which is inclined with respect 
to the orbital plane. When the neutron star crosses the disk,
the mass accretion rate onto the neutron star, therefore the  X-ray flux,
increases temporarily. Transient $\sim 18$ sec oscillations have appeared 
during the secondary flare (In 't Zand, Baykal $\&$ Strohmayer 1998). These 
oscillations may be interpreted as Keplerian motion of an accretion 
disk near the magnetospheric radius. Due to the long spin period  
the co-rotation radius is much larger than the magnetospheric radius 
corresponding to the magnetic field of 2.1 $\times$ 10$^{12}$ Gauss implied 
by a cyclotron feature in the X-ray spectrum (Cusumano et al., 1998). 
Therefore 4U 1907+09 is not likely to be spinning near equilibrium,
unlike other accretion powered X-ray pulsars.
The 18 second quasi periodic oscillation at the flare
 suggests the formation of transient accretion disks from the wind accretion 
(In 't Zand, Baykal $\&$ Strohmayer 1998).
Another interesting feature of the source
is the sudden decrease in X-ray intensity by a factor of 
$\sim$ 10$^{2}$,
during time intervals ranging from a few minutes to $\sim$ 1.5 hour 
(In 't Zand, Strohmayer $\&$ Baykal 1997). The spectra at the dipping 
activity and outside the dip periods were similar 
with no indication of large 
changes in the column density of cold circumstellar 
matter (i.e. N$_{H}$ remains below 10$^{23}$ cm$^{-2}$). It is 
suggested that the mass accretion rate ceased 
due to the inhomogeneous spherical wind from the companion.

In this work, we have investigated the stability of the spin down rate. 
This source has shown spin down rate changes less than $\sim$ 8 $\%$
within 12 years (In 't Zand, Strohmayer $\&$ Baykal 1998).
 Using the archival RXTE observations, we measured 
three new pulse periods covering a time span of over 
2 years in addition to the previous four pulse period measurements. 
With $\sim 10^{3}-10^{4}$ sec observations separated by 
intervals of the order of a month 
 we have been able to connect the pulses 
in phase and to construct the timing solution extending 
over a year. The residuals of pulse arrival times yielded a very low 
noise strength. Our findings imply that the source 
has a very stable spin down rate even over short time intervals, in contrast 
to the noise seen in other HMXRBs.

\section{Observations and Results} 

The observations used in this work are listed in Table 1.
The results presented here are based on data collected
with the Proportional Counter Array (PCA, Jahoda et al., 1996).
The PCA instrument consists of an array 
of 5 proportional counters operating in the 2-60 keV energy range, with 
a total effective area of approximately 7000 cm$^{2}$ and a field of view 
of $\sim 1^{\circ}$ full width half half maximum.

Background light curves were generated using
the background estimator models based on the rate of very large events (VLE),
spacecraft activation and cosmic X-ray emission with the standard PCA
analysis tools (ftools) 
and were subtracted from the source light curve obtained from
the event data. The background subtracted light
curves were corrected with respect to the barycenter of the solar system. 
Using the binary orbital parameters of 4U 1907+09 from RXTE observations 
(In 't Zand, Baykal $\&$ Strohmayer 1998), the light curves 
are also corrected for binary motion of 4U 1907+09 (see Table 3).
From the long archival data string outside the intensity dips,
pulse periods for 4U 1907+09 were found by folding the time series 
on statistically independent trial periods (Leahy et al. 1983).
Master pulses were constructed from these observations
by folding the data on the period giving the maximum $\chi^2$.
The master pulses were arranged in 20 phase
bins and represented by their Fourier harmonics (Deeter \& Boynton 1985) and
cross-correlated with the harmonic representation of average pulse profiles
from each observation. The pulse arrival times are obtained from the 
cross-correlation analysis. We have measured three new pulse periods from 
the longer observations. These are presented in 
Figure 1 and listed in Table 2.  
We have found that the rate of change of the pulse period of 4U 1907+09
is stable. Therefore we have been able to connect all
pulse arrival times in phase over a 383 day time span. 
The pulse arrival 
times are fitted to the quadratic polynomial  
\begin{equation}
\delta \phi = \phi_{o} + \delta \nu (t-t_{o})
+ \frac{1}{2} \dot \nu (t-t_{o})^{2}
\end{equation}
where $\delta \phi $ is the pulse phase offset deduced from the pulse
timing analysis, $t_{o}$ is the mid-time of the observation, $\phi_{o}$ is
the phase offset at t$_{o}$, $\delta \nu$ is the deviation from the mean
pulse frequency (or additive correction to the pulse frequency), and $\dot
\nu $ is the pulse frequency derivative of the source. 
The pulse arrival times (pulse cycles) and 
the residuals of the fit after the removal of the quadratic polynomial
are presented in the Figure 2 and Figure 3 respectively.
Table 3 presents the timing solution of 4U 1907+09.
The pulse frequency derivative $\dot \nu = (-3.188 \pm  0.006) \times
10^{-14}$ Hz s$^{-1}$ is measured from the pulse arrival times
obtained in a sequence of 19 observations
spread over 383 days. This value is consistent 
within 10 $\%$ with the long term 
value obtained from the data displayed in Figure 1,
 $\dot \nu$ = (-3.54 $\pm  0.02) \times 10^{-14}$ Hz s$^{-1}$.
The residuals of the fit give a random walk noise strengths at
$T_{observation} \sim 383$ days,
S$\approx (2\pi)^{2}<\delta \phi ^{2}>/T^{3}_{observation}
\approx (2\pi)^{2}<\delta \nu ^{2}>/T_{observation}
\sim 2 \times 10^{-20}$ rad$^{2}$ sec$^{-3}$,
where $<\delta \phi ^{2}>$ and $<\delta \nu ^{2}>$ are the normalized
variances of pulse arrival times and residual pulse frequencies
(see Cordes 1980 for further definitions of noise strength).
This value is 4 decades lower than that of Vela X-1  (Bildsten et al., 1997)
and it is only a factor 5 greater than that of the LMXRB pulsar 4U 1626-67
(Chakrabarty et al., 1997). The noise strength of 4U 1626-67 was considered 
the smallest ever measured for an accretion powered X-ray source. 
This noise strength is indeed very low for a HMXRB pulsar. 
The stable spin down rate over the 
15 years and the low level of noise strength is a unique property of this 
source among the HMXRBs. The spin down rate of 4U 1907+09 
is only a factor four greater than that of the AXP source 1E 2259+586. 
Furthermore the long term noise strength is one order of magnitude lower than  
the AXP 1E 2259+586 (Baykal $\&$ Swank 1986).
The quiet and persistent spin down rate of 4U 1907+09 shows that 
an accreting pulsar can spin down quietly, 
for extended periods.  
For the AXPs as well as 1E 2259+586, the existence of long 
epochs of spin down has been interpreted as evidence that these sources 
are isolated pulsars in dipole spin down in which case the large spin down 
rates and periods would indicate large 
(10$^{14}$-10$^{15}$ Gauss) magnetic fields (Thompson $\&$ Duncan 1993). 
The existence of known accreting sources with quiet and persistent 
spin down, as observed from 4U 1626-67, and now 4U 1907+09 shows that 
quiet spin down does not necessarily imply that the source is 
not accreting.

{\bf Acknowledgements~:}\\
We thank Dr. Jean Swank for the helpful comments. \\

\begin{table} 
\caption{Observation List for 4U 1907+09}
\label{Pri}
\[
\begin{tabular}{c c c}\hline
Time of Observation & Exposure      \\
 day/month/year      &  sec            \\ \hline \hline
25/11/1996 & 9163 \\
19-27/12/1996 & 35102 \\
29/01/1997 & 849 \\
19/03/1997 & 7430 \\
29/04/1997 & 13908 \\
26/05/1997 & 8352 \\
18/06/1997 & 11695 \\
17/07/1997 & 724 \\
24/08/1997 & 6976 \\
23/09/1997 & 5811 \\
18/10/1997 & 7913 \\
17/11/1997 & 7787 \\
14/12/1997 & 645 \\
26-29/07/1998 & 33211 \\
18/09-01/10/1998 & 175382 \\
\hline
\end{tabular}
\]
\end{table}

\newpage

\begin{table}
\caption{RXTE Pulse Period Measurements of 4U 1907+09
   }
\label{Pri}
\[
\begin{tabular}{ c c c }  \hline
Epoch(MJD)  & Pulse Period (sec)   & Ref.      \\ \hline
45576 &      437.483 $\pm$ 0.004  & Makishima et al., 1984 \\   
45850 &      437.649 $\pm$ 0.019 & Cook $\&$ Page 1987 \\
48156.6 &    439.19 $\pm$ 0.02  & Mihara 1995 \\
50134   &    440.341 $\pm$ 0.014 & in't Zand et al., 1998 \\
50440.4  &440.4877$\pm$0.0085  & This work  \\
51021.9  &440.7045$\pm$0.0032  & This work   \\
51080.9  &440.7598$\pm$0.0010  & This work  \\ \hline

\end{tabular}
\]
\end{table}

\begin{table}
\caption{Timing Solution of 4U 1907+09 for RXTE Observations$^{a}:$
   }
\label{Pri}
\[
\begin{tabular}{c|c}  \hline
Orbital Epoch (MJD) & 50134.76(6)$^{b}$ \\
P$_{orb}$ (days)    & 8.3753(1)$^{b}$ \\
a$_{x}$ sin i (lt-sec) & 83(2)$^{b}$  \\
e                       & 0.28(4)$^{b}$  \\
w   & 330(7)$^{b}$ \\
Epoch(MJD)  & 50559.5011(3)      \\
Pulse Period (sec)  & 440.5738(2) \\
Pulse Period Derivative  (s s$^{-1}$) & 6.18(1)$\times$ 10$^{-9}$ \\ 
Pulse Freq. Derivative (Hz s$^{-1}$)  &
-3.188(6)$\times 10^{-14}$ \\ \hline
\end{tabular}
\]
$^{a}$ Confidence intervals are quoted at the 1 $\sigma $ level. \\
$^{b}$ Orbital parameters are taken from in't Zand et al., 1998 et al., (1997).
P$_{orb}$=orbital period, a$_{x}$ sin i=projected semimajor axis,
e=eccentricity, w=longitude of periastron.\\
\end{table}

{\Large{\bf Figure Caption}}\\

{\bf Fig. 1}~~Pulse period history of 4U 1907+09.\\

{\bf Fig. 2}~~Pulse phase (Pulse Cycles) of 4U 1907+09 
with respect to the constant pulse period of 440.5738 sec.

{\bf Fig. 3}~~Pulse phase residuals of 4U 1907+09 
with respect to the constant pulse period of 440.5738 sec after
the derivative of pulse period 6.18 $\times$ 10$^{-9}$ s s$^{-1}$ is removed.

\clearpage
\begin{figure}
\plotone{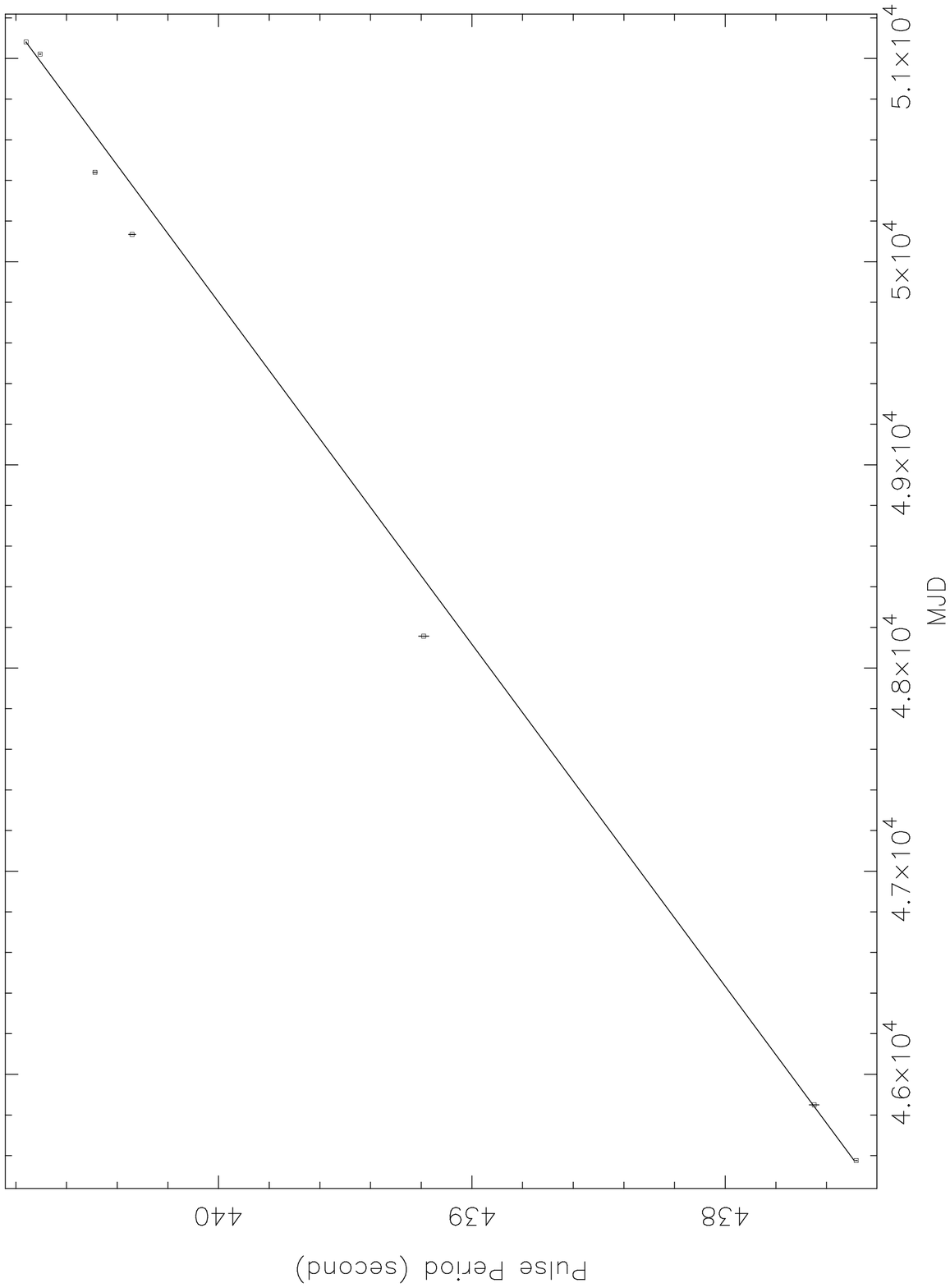}
\end{figure}

\clearpage
\begin{figure}
\plotone{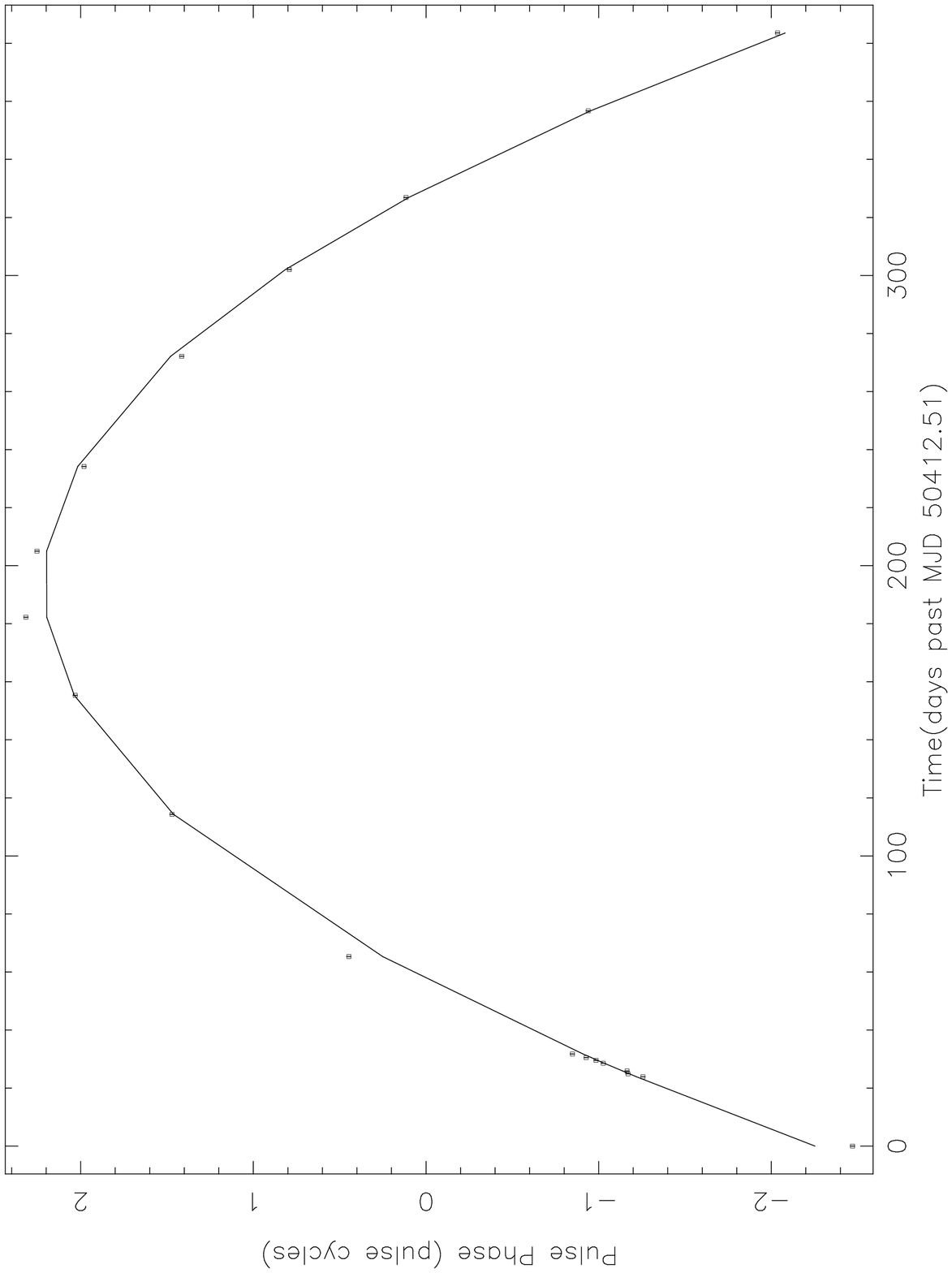}
\end{figure}

\newpage
\clearpage
\begin{figure}
\plotone{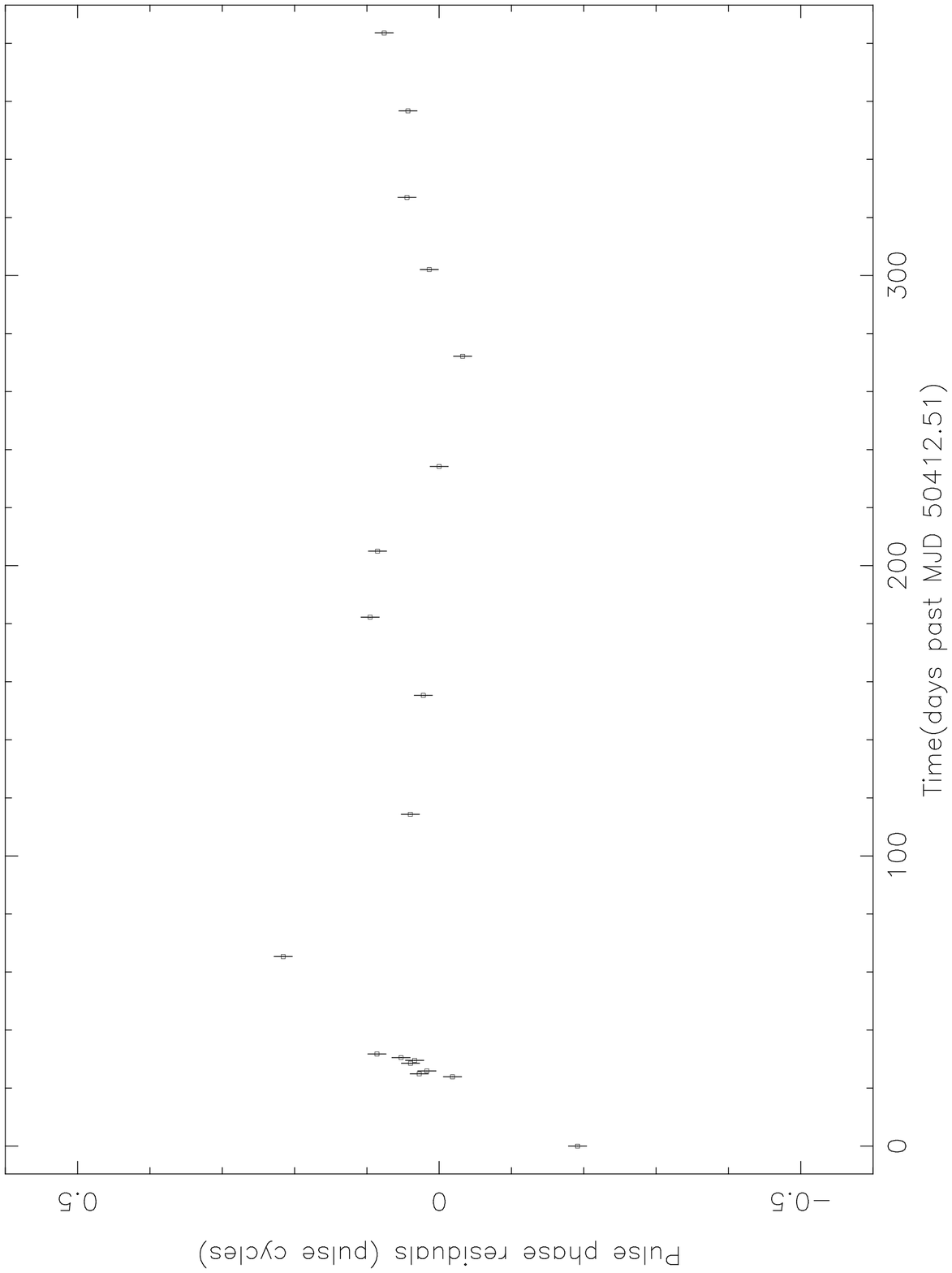}
\end{figure}

\end{document}